# The evolution of microRNA-regulation in duplicated genes facilitates expression divergence.

Yonatan Bilu, Department of Molecular Genetics, Weizmann Institute of Science, Rehovot 76100, Israel.

**Abstract**

*Background:* The evolution of microRNA regulation in metazoans is a mysterious process: MicroRNA sequences are highly conserved among distal organisms, but on the other hand, there is no evident conservation of their targets.

*Results:* We study this extensive rewiring of the microRNA regulatory network by analyzing the evolutionary trajectories of duplicated genes in D. melanogatser. We find that in general microRNA-targeted genes tend to avoid gene duplication. However, in cases where gene duplication is evident, we find that the gene that displays high divergence from the ancestral gene at the sequence level is also likely to be associated in an opposing manner with the microRNA regulatory system – if the ancestral gene is a miRNA target then the divergent gene tends not to be, and vice versa.

*Conclusions:* This suggests that miRNAs not only have a role in conferring expression robustness, as was suggested by previous works, but are also an accessible tool in evolving expression divergence.

**Background**

MicroRNAs (miRNA) are short RNA molecules that inhibit protein synthesis by targeting mRNA transcripts [1]. Intriguingly, while miRNAs typically exhibit very high sequence conservation even among evolutionarily distant species, their mRNA targets are specific to each species, with as little overlap as can be expected by chance [2-5]. This portrays the evolution of regulation by miRNAs as a mysterious process, in which the regulatory elements are kept under strong stabilizing selection, while the elements which they regulate are free to diverge.

The lack of target conservation may indicate a high turnover of *cis*-elements in the 3' UTR of mRNA transcripts, similar to what has been suggested for transcription factor binding sites (TFBSs) in gene promoter sequences [6-12]. Several works have argued that the latter may, in fact, have little effect on phenotype, as changes can be compensatory, and even when they are not, mRNA expression patterns may often change by way of neutral drift [13, 14]. Could the same be true for miRNA regulation?

Notably, evolution of miRNA regulation is reminiscent, yet ultimately quite different, from regulation mediated by transcription-factors [15]. First, while transcription factors are more conserved than other protein coding genes, they are not as highly conserved as miRNAs. Second, while conservation of TFBSs is lower then might be expected [16-18], it is still relatively higher than for miRNA binding sites, especially among highly conserved transcription factors. Third, transcription-factor binding sites in eukaryotes tend to be short and fuzzy [9, 19]. Binding of miRNA is also thought to be facilitated by short (7bp) sequences [20-22], but less tolerant of fuzziness, perhaps due to a smaller amount of structural cues. Finally, while fitness-neutral changes are possible for both types of regulation, changes in transcription regulation can be more readily compensated for at intermediate levels, which are not relevant for post-transcriptional regulation.

The basic dogma for genetic diversity is that genes undergo duplication, and then one of the copies is free to evolve and explore new functional roles [23-25]. While this view may be simplistic, it highlights the study of gene duplicates as a key tool in understanding evolutionary processes. Indeed, such studies are the basis for identification of whole genome duplication events [23], and cases of *subfunctionalization* and *neofunctionalization* [26, 27].

In this work we study the evolution of miRNA regulation as it pertains to gene duplicates in *Drosophila melanogaster*. We find that miRNA regulation is relatively rare among gene duplicates, possibly due to an indirect selective bias against duplication of miRNA targeted genes. Nonetheless, close examination of miRNA targeting in A. gambiae and D. melanogaster suggests miRNA regulation has an important role in facilitating the exploration of novel functionality. This "exploratory" role may account for the lack of conservation in miRNA targets, and can be seen as complementary to the suggested role of miRNAs in conferring expression robustness [28-30].

**Results**

*miRNA target conservation and 3' UTR length*

As previously reported, miRNA target conservation is close to what may be expected at random [15]. Specifically, we compared the targets predicted by miRanda [31] for 14 miRNAs conserved between H. sapiens and D. melanogaster. For each miRNA, we asked how many of the D. melanogaster orthologs of its H. sapiens targets are targeted by it in D. melanogaster (Figure 1). For 9 out of the 14 this number is almost exactly what is expected at random, for 3 it is higher, and for 2 it is, in fact, lower. This lack of conservation is especially striking, since miRanda makes use of sequence conservation (albeit in D. pseudoobscura and A. gambiae) to generate predictions.

Could this seemingly complete turnover be a result of compensatory changes, where miRNA-regulated transcripts shift among targeting miRNAs? A statistical analysis suggests that this might indeed be the case to some extent, as D. melanogaster genes whose H. sapiens orthologs are miRNA targets, are significantly more likely to be miRNA targets themselves (Table 1). In other words, while the targets of specific miRNAs are not conserved between H. sapiens and D. melanogaster, the property of simply being part of the miRNA-regulatory network is.

Notably, this conservation, while statistically significant, is not striking (31.22% vs. 22.56% expected). One possibility is that the current state-of-the-art in predicting miRNA targets computationally still leads to many false predictions (both positive and negative), masking an otherwise stronger trend. On the other hand, as noted above, a bias towards conserved sites is expected. Moreover, target prediction algorithms will tend to classify (perhaps rightly) mRNAs with long 3' UTRs as miRNA targets, if only for the reason that there are more sub-sequence candidates for miRNA-binding in the long UTRs. Thus, the observation that genes tend to "remain" miRNA targets, may simply represent a conserved enrichment for long 3' UTRs in these genes.

Correcting for UTR length is not trivial, since, as shown by Stark et al. [29], the number of predicted binding sites does not scale linearly with 3' UTR length. Stark et al. observed that longer 3' UTRs have more binding sites per nucleotide than shorter ones. This is also the case for PicTar predictions (Figure 2A), for which this binding-site density seems to increase in 3' UTRs of 800bp and longer. Yet, there is a caveat here - while there are more binding sites, these binding sites belong (on average) to less miRNAs (Figure 2B). Moreover, basing this analysis on miRanda target predictions suggests an opposite phenomenon - 3' UTRs of 500bp and longer actually have a lower miRNA binding site density than shorter ones (and fewer targeting miRNAs, per nucleotide; Figures 2C and 2D).

Interestingly, this dichotomy between long and short 3' UTRs is also apparent at the level of the miRNAs themselves. We observed that the number of transcripts that a miRNA targets is negatively correlated with the fraction of its targets that have a long 3' UTR (> 800; Spearman r = -0.46, p=$3.2 \times 10^{-5}$; Figure 3).

In summary, while there is some indication for compensatory changes in miRNA-regulation, the current state of the art in miRNA targets prediction is such that it is difficult to assess if this is indeed the main reason for target turnover. Thus, to study how genes become miRNA targets during evolution, or cease to be ones, there is a need to focus on relatively discernable events. Towards this end, we now turn our attention to paralogous genes, and aim to identify trends of acquisition and loss of targeting miRNA following gene duplication.

*Paralogs avoid miRNA regulation*

As with miRNA target prediction, identifying paralogous genes computationally can be done in several ways. The most straightforward is to consider the relevant genome, and define two genes as paralogs if their sequence similarity exceeds some threshold (see e.g. [32]). Alternatively, when sequenced genomes of related organisms are available, they can serve as a reference point for more accurate predictions. For example, if a single gene in one organism has two very similar copies in the other, it is likely that they are paralogs (either as a result of gene duplication that occurred after the species diverged, or as a result of prior gene duplication, after which one copy was lost in the reference organism).

Two possible scenarios for the evolution of miRNA regulation following gene duplication present themselves. In the "neutral turnover" scenario, similarly to what is suggested for transcriptional regulation, changes in miRNA targeting are mainly a result of a phenotypic-neutral turnover, perhaps due to compensatory changes in which the identity of the targeting miRNA changes, but repression by some miRNA is maintained. Alternatively, in the "expression differentiation" scenario, miRNA regulation is a prevalent method in differentiating protein expression following gene duplication, perhaps leading to novel functions. The latter seems an appealing mechanism, whereby the translation of one gene copy is inhibited, and therefore the duplication does not change protein content dramatically, allowing the duplicated gene to explore new roles.

To discern between the two scenarios, we asked whether genes with paralogs are as likely a target of miRNA regulation, as may be expected from a neutral turnover of targets, or are enriched for miRNA targets, as may be expected from the "expression differentiation" scenario. Surprisingly, using several methods for defining paralogs (see Methods), and using both PicTar and miRanda algorithms for target prediction, we consistently see that duplicated genes actually tend to avoid miRNA regulation (Table 2).

One possibility for this is that retention of gene duplicates is biased for certain gene properties, which make miRNA repression less likely. For example, at the first stage after gene duplication the transcript level of the gene will probably double. Hence, genes which are likely to be maintained after duplication are those for which a high copy number is beneficial, or at least non-deleterious. These genes may be a-priori less likely miRNA targets. In other words, although there might be a neutral turnover of miRNA control (or even a weak trend towards expression divergence), the effect will be masked by a selection bias for duplicates which are not microRNA targets, resulting in a low number of miRNA among genes with paralogs.

To test this, we asked whether genes whose expression level varies over a large range of values, and thus, conceivably, their duplication impact will be less severe (since the system already deals with dosage variability), are enriched for miRNA targets. We find that indeed there is some negative correlation between a gene being a miRNA target, and the variance of the gene's expression (during development), but this correlation, although statistically significant ($p=5.16 \times 10^{-8}$) is very weak ($r=-0.08$; Spearman rank correlation).

*Regulation at multiple levels*

Regulation of protein abundance is done at multiple levels, among which miRNA regulation is just one. An interesting question which will be instrumental to the analysis the follows, is whether there is a trade-off of regulation among these levels, or a correlation among them. For example, if there is a trade-off, then genes which are tightly regulated by transcription factors are less likely to be miRNA targets, since their repression is already achieved at the transcriptional level. Alternatively, being "highly regulated" is a property of gene that will manifest at all levels of regulation.

We start by analyzing the promoter regions of microRNA targets, which plausibly reflect the extent of a gene's transcriptional regulation. We observe a positive correlation between the number of targeting miRNAs and both the conservation of the gene's promoter region, and the number of predicted TFBSs therein. These two attributes may reflect a high level of transcriptional regulation, as was suggested for S. cerevisea [33]. However, as with expression variance, these correlations, although statistically significant ($p=3.12 \times 10^{-9}$ for promoter conservation, 0.007 for TFBS content) are very weak ($r=0.05, 0.02$; Spearman rank correlation).

Nonetheless, they may suggest that miRNA targets are genes whose expression is highly regulated, both transcriptionally and post-transcriptionally, and hence, their chance duplication, which dramatically disrupts their expression level, is deleterious. If this is indeed true, then it might be expected that miRNA targets whose duplication does get fixated will tend to be those whose expression is under loose control, and free to evolve. At first glance, this indeed seems to be reflected at the transcriptional level, as paralogous pairs which are both miRNA targets have a strikingly lower similarity in TFBSs content than pairs where neither is a target (Table 3).

Interestingly, we find that the promoter regions of duplicated genes are much less conserved than those which are not (Table 4), suggesting that either the regulation of duplicated genes tends to diverge, as each copy assumes a somewhat different role, or that retention of gene duplicates tends to occur when strict regulation is less important and free to diverge. To the extent that the latter is true, this is consistent with the finding above, suggesting that regulatory rewiring occurs where there is less selection against it (as was also suggested for S. cerevisae [33]).

However, examining the genes in question, we see that difference between targeted and non-targeted paralogous pairs is strongly biased by the numerous clusters of histone-coding genes (His1, His2A, His2B, His3 and His4; whose cluster is duplicated over twenty times in the D. melanogaster genome). When the highly conserved promoters of these genes are excluded, the difference is apparent only for BLAST-based paralogs.

Another layer of post-transcriptional regulation is the alternative splicing of transcripts. Thus, if a high level of regulation is a property that tends to be manifested at multiple levels, it is expected that genes with multiple splice-forms will be enriched for miRNA targets. Indeed, we observe that the number of targeting miRNAs is correlated with the number of alternative splice forms ($r=0.27$, $p=4.78 \times 10^{-233}$ for PicTar; $r=0.05$, $p=1.18 \times 10^{-8}$ for miRanda; Spearman rank correlation). Consistently, we find that genes with multiple splice-forms are less prone to be duplicated (Table 5).

*Loss of miRNA regulation as a mean for expression divergence*
A second explanation for the under-representation of microRNA targets among duplicated genes may be that miRNA-mediated expression diversification following duplication does not occur mainly due to acquisition of miRNA binding sites, but rather via their loss. If this is indeed a main trend in the evolution of miRNA regulation, it is expected that when paralogy is defined via a reference to another organism, the single copy in this reference organism will tend to be a miRNA target. In other words, in this scenario there is a bias towards duplication of miRNAs, and following duplication, one of the copies (or both) will cease to be target.

However, if anything, we see an opposite phenomenon. Orthologs of D. melanogaster genes in D. pseudoobscura and A. gambiae which have only one ortholog in D. melanogaster are enriched for miRNA targets (Figure 4; miRanda target predictions). This further supports the possibility that duplication of miRNA target genes is less likely to become fixed during evolution.

*Expression differentiation in duplicated genes*

According to the "expression differentiation" scenario, following duplication one copy maintains the original role, while the other is inhibited, and relatively free to explore new ones. This functional exploration is likely to be evident from both the expression pattern of the genes, and their sequence. Hence, when using a reference organism to define paralogy, we discern between the "close ortholog" of the reference gene, which is assumed to remain under stabilizing selection, and the "remote ortholog" which is relatively free to diverge. If miRNA regulation tends to evolve as a result of "expression differentiation" then it is expected that acquisition of miRNA regulation (or its loss) will tend to occur in the "remote ortholog". If, on the other hand, miRNA targeting tends to occur as a neutral turnover, then no correlation between the two properties is expected.

For this analysis, we focused on A. gambiae as the reference organism, since it is more remote, and thus the associated paralogous pairs had more time to diverge. Accordingly, we consider only miRanda target predictions, as PicTar predictions are not available for A. gambiae. Interestingly, Table 6 supports the "expression differentiation" scenario mentioned above, where miRNA targeting of the "remote ortholog" tends to be opposite to that of the reference gene. Moreover, this trend is decoupled from the bias toward long 3' UTRs – "close orthologs" tend to have longer 3' UTR, regardless of whether or not the reference gene is a miRNA target.

**Discussion and Conlusions**

Conservation of miRNA sequences among a wide range of organisms stands in sharp contrast to a nearly complete turnover of their targeted mRNA transcripts. From a systems point of view, this high turnover rate may be a byproduct of the high level of backup suggested for the miRNA regulatory system [28, 34]. For example, in the bristle scutellar development system, five miRNAs co-target most of the associated gene transcripts [28]. Similarly, at a more global level, we analyzed the clustering coefficient of the graph defined by the miRNA co-targeting (unpublished work), and found it to be significantly higher than expected at random (p-value < $10^{-14}$). Hence, mutation in the 3' UTR of a gene which leads to a loss of a miRNA binding site may be compensated for by binding sites for other miRNAs. Indeed, as reported by Stark et al. [29], in D. melanogaster multiple binding sites in the 3' UTR of a gene tend to be associated with different miRNAs.

What distinguishes miRNA-targeted genes from the rest of the genes? Analyzing the GO [35] categories of such genes, and correcting for multiple hypothesis testing, we could not identify a convincingly strong enrichment for a particular function. However, Cui et al. [7] have found that the expression of miRNA-regulated genes tends to be evolutionarily conserved. They concluded from this that miRNAs can affect gene expression by reducing stochastic noise, buffer cross-species variation and constrain the evolution of gene expression variation. Put differently, miRNAs tend to target genes for which there is selective pressure against expression variation. Thus, it is not surprising that we find that miRNA-regulated genes are less likely to undergo duplication, as the resulting disruption of expression patterns is likely to be deleterious.

Interestingly, our comparative study of miRNA targets in D. melanogatser and A. gambiae suggests that those miRNA targets that did undergo duplication tend to display an evolutionary pattern that is suggestive of neofunctionalization facilitated by divergence at multiple levels: In one gene the sequence and the regulatory status are conserved, while in the other gene the sequence diverges, and the regulatory status changes – if the ancestral gene is a miRNA target the divergent gene tends not to be one, and vice versa. Thus, it is possible that in addition to having a role in stabilizing expression [28-30], miRNA regulation, because of its high plasticity, may actually facilitate expression divergence in some cases.

Expression-facilitated neofunctionalization was also observed in yeast by Tirosh and Barkai [27], who identified 43 duplicate pairs where the expression of one gene is markedly more divergent than the other. However, in contrast to the analysis here, they found only a weak correlation between expression divergence and sequence divergence. This may suggest that in higher eukaryotes sequence and expression divergence are more correlated, possibly due to stronger constraints on genes under stabilizing selection, or because the evolution of miRNA regulation is inherently different from that of transcription regulation. The latter is consistent with our finding that paralogous pairs in which both genes are miRNA targets tend to be have low TFBS-content similarity of their promter regions.

In summary, while the miRNA regulatory system displays extensive rewiring, we this work suggests that the identity of the genes associated with this network is conserved to some extent. The expression of these miRNA-regulated genes are suggested to be under stabilizing selection, both by previous work [7], and by the observation that they are less likely to be duplicated. However, in cases where duplication is evident, we find that miRNA regulation may be a common way for achieving expression differentiation between the two copies. This portrays microRNA regulation as having a twofaced role in evolution – conferring expression robustness [28, 30], but also facilitating expression divergence.

**Methods**

  *1. Datasets*

Genomic sequences of *D.* melanogaster were downloaded from FlyBase [36] (version 4.2; http://flybase.bio.indiana.edu/), as well as protein sequences, locations of protein-coding genes and data on alternative splice forms. Lengths of 3' UTR length was downloaded from UCSC's Genome Browser ([37]; http://genome.ucsc.edu).

MicroRNAs and their miRanda-predicted targets were downloaded from miRBase ([31, 38]; http://microrna.sanger.ac.uk). PicTar targets predictions were downloaded from the 2006 dataset ([5]; http://pictar.bio.nyu.edu).

TF binding sites motifs were taken from the list compiled by Elemento and Tavazoie [39] for the D. melanogaster network (371 highest scoring k-mers). Conservation score for promoter regions were taken from UCSC's Genome Browser ([37]; http://genome.ucsc.edu).

Expression data was taken from the Stanford Microarray Database ([40]; http://genome-www5.stanford.edu; D. melanogaster development series).

  *2. Conserved miRNA-targets Expectation*

The expected number of conserved targets for each miRNA was computed according to a binomial distribution. Suppose miRNA *r* targets *m* mRNA transcripts in human which have well defined orthologs (see *Paralogs*, below), and that all in all there are *n* genes in D. melanogaster which have a well defined ortholog in human and are targeted by miRNA. Denote the number of targets for miRNA *r* in D. melanogaster by *k*. Then the number of targets for *r* in D. melanogaster which is expected to coincide by chance with the orthologs of its human targets is: $\frac{k \bullet m}{n}$.

The expected overlap between miRNA-targets in D. melanogaster and orthologs of miRNA-targets in human was computed similarly, with *n* taken to be the total number of genes in D. melanogaster which a have a well defined ortholog in human.

3. Paralogs

Genes were identified as paralogs based on three different methodologies. InParanoid [41] was used to determine orthology between D. melanogaster and D. pseudoobscura, and multiple D. melangosater genes orthologous to a single D. pseudoobscura gene were considered paralogs. Similarly, Ensembl-BioMart ([42]; http://www.ensembl.org/biomart/martview) orthology predictions were used to determine paralogy based on orthology with A. gambiae.

InParanoid was also used to identify H. sapiens genes with their D. melanogaster orthologs. Only 1-to-1 correspondence was considered.

All-against-all Blast comparisons were used to defined intra-genome paralogy. Two genes were defined as paralogs if their sequence similarity E-scores are both at most $10^{-10}$, and the ratio of their length is at most 1.5. This latter condition deals with the problem of very short proteins that are highly similar to a short fragment of a very long protein.

4. TFBS similarity

The 2000bp upstream of each gene in D. melanogaster were scanned for exact matches to the putative TFBS sequences defined by Elemento and Tavazoie [39]. The TFBS-content similarity score for a pairs of genes is defines by the fraction of TFBS sequences appearing in both their promoters.

Namely, if the list of putative TFBSs appearing in the promoter region of gene1 is *L1*, and that of putative TFBSs appearing in the promoter region of gene2 is *L2* then their similarity score is defined as $\frac{|L1 \cap L2|}{|L1 \cup L2|}$.

5. Promoter conservation score

Promoter conservation scores are based on the conservation score defined in UCSC's Genome Browser for each position in the 2000bp upstream of each gene. All scores exceeding some threshold were added up to define the conservation score of the entire promoter region. We report here results for a threshold of 0.9, but other thresholds in the range 0.5-0.9 yielded similar results.

### 6. *Remote and close orthologs*

Pairs of genes in D. melanogaster with a 2-to-1 orthology relation to an A. gambiae gene were labeled as "remote" and "close" ortholog based on sequence identity of the coded proteins to their A. gambiae counterpart. If the difference in identity percentage exceeded 10%, the more similar one was labeled as "close" and the other as "remote". Otherwise, both were considered to be similarly distant from their A. gambiae ortholog. Out of a total of 507 paralog-pairs 463 displayed such an identity difference and were included in the analysis.

### 7. *Co-targeting graph*

The co-targets graph is defined over miRNAs, based on their common targets. This graph is a weighted graph, where the weight of an edge between two nodes is the number of their common targets divided by the total number of their targets. For example, suppose miRNA1 targets geness g1, g2 and g3; and miRNA2 targets genes g2, g3 and g4. Then in the overlap graph, there is an edge between miRNA1 and miRNA2 with weight 2/4=0.5 (they target two common genes, and a together target a total of four genes).

The clustering coefficient of a node in a simple graph is the number of edges the number of edges among its neighbors divided by the maximal possible number of edges among its neighbors [43]. Specifically, if a node has $k$ neighbors with $n$ edges among them, its clustering coefficient is *$2n/(k(k-1))$*. The clustering coefficient of a graph is the mean clustering coefficient of its nodes. Nodes of degree less than two are omitted from the analysis (their clustering coefficient is undefined). In a weighted graph, with edge weights in [0,1], the clustering coefficient is defined similarly with the modification that the $n$ above is replaced by the sum of weights of the edges among the node's neighbors.

The significance of the clustering coefficient is estimated by generating random graphs with the same degree distribution (using the "edge swapping" algorithm), and computing the mean and variance of their clustering coefficient. Assuming a normal distribution for these values, the p-value for the observed clustering coefficient is deduced.

**List of abbreviations**

miRNA – MicroRNA

mRNA – Messenger RNA

TF – Transcription Factor

TFBS - Transcription Factor Binding site

UTR – Untranslated region

**Acknowledgements**

This work was funded by the UniNet EC NEST consortium contract number 12990. I thank Itay Tirosh and Naama Barkai for helpful discussions.

Tables

|  | Fraction of targeted orthologs | Expected fractions of targeted genes | p-value |
|---|---|---|---|
| miRanda predictions | 31.98 % | 22.56 % | $\ll 10^{-100}$ |
| Pictar predictions | 26.05 % | 20.85 % | $4.81 \times 10^{-10}$ |

**Table 1**: Fraction of D. melanogatser orthologs of H. sapiens miRNA-targeted genes, which are themselves miRNA targets.

| *Prediction algorithm* | miRanda | | PicTar | |
|---|---|---|---|---|
| *Paralogy method* | 1:1 orthologs | paralogs | 1:1 orthologs | paralogs |
| Blast matches | 0.2428 | 0.1404 | 0.2307 | 0.0815 |
| Comparison to A. gambiae | 0.3110 | 0.2133 | 0.3291 | 0.1596 |
| Comparison to D. pseudoobscura | 0.2632 | 0.1748 | 0.2557 | 0.0728 |
| all genes | 0.2386 | | 0.2238 | |

**Table 2:** Fraction of genes with and without paralogs, which are miRNA targets. The table lists values for different methods for defining paralogy, and for different miRNA target prediction algorithms.

|  | Blast matches | Comparison to D. pseudoobscura | Comparison to A. gambiae |
|---|---|---|---|
| Both are miRNA targets | 0.1627 (52) | 0.3854 (4) | 0.1339 (154) |
| One is, other isn't | 0.2252 (105) | 0.3040 (38) | 0.1096 (978) |
| At least one is | 0.2031 (147) | 0.3117 (42) | 0.1129 (1132) |
| Neither is | 0.8543 (2840) | 0.5505 (604) | 0.2714 (7488) |
| Neither is, His genes excluded | 0.5150 (480) | 0.2183 (260) | 0.1386 (6196) |

**Table 3:** Mean similarity of transcription factor binding sites content, for paralogous pairs, as a function of whether they are miRNA targets. Numbers in parenthesis indicate the number of genes in the category.

|  | Genes with paralogs | Genes with no paralogs | His genes with paralogs |
|---|---|---|---|
| *Paralogy method* | | | |
| Blast matches | 557.6265 | 602.6264 | 776.5162 |
| Comparison to A. gambiae | 509.7652 | 617.4745 | 716.0834 |
| Comparison to D. pseudoobscura | 585.6993 | 632.381 | 714.472 |

**Table 4:** Conservation score of 2000bp promoter regions for genes with and without paralogs.

|  | Genes with paralogs | Genes with no paralogs | His genes with paralogs |
|---|---|---|---|
| *Paralogy method* | | | |
| Blast matches | 0.22 | 0.12 | 0 |
| Comparison to A. gambiae | 0.31 | 0.18 | 0.02 |
| Comparison to D. pseudoobscura | 0.23 | 0.09 | 0.14 |

**Table 5:** Fraction of genes with multiple isoforms.

| Anopheles ortholog | Close ortholog | Distant ortholog | Overall |
|---|---|---|---|
| …is a miRNA target | 0.1882 (423.44) | 0.2941 (226.67) | 0.2645 |
| …is not a miRNA target | 0.2988 (340.29) | 0.1992 (247.62) | 0.2575 |

**Table 6:** Probability that a D. melanogatser gene is a miRNA target depending on whether it is the "close" or "distant" ortholog of an A. gambiae gene, and on whether or not this ortholog is a miRNA target. Numbers in parenthesis denote the mean 3' UTR length for each set of genes. Data is given for a total of 463 paralogous pairs.

## Figures

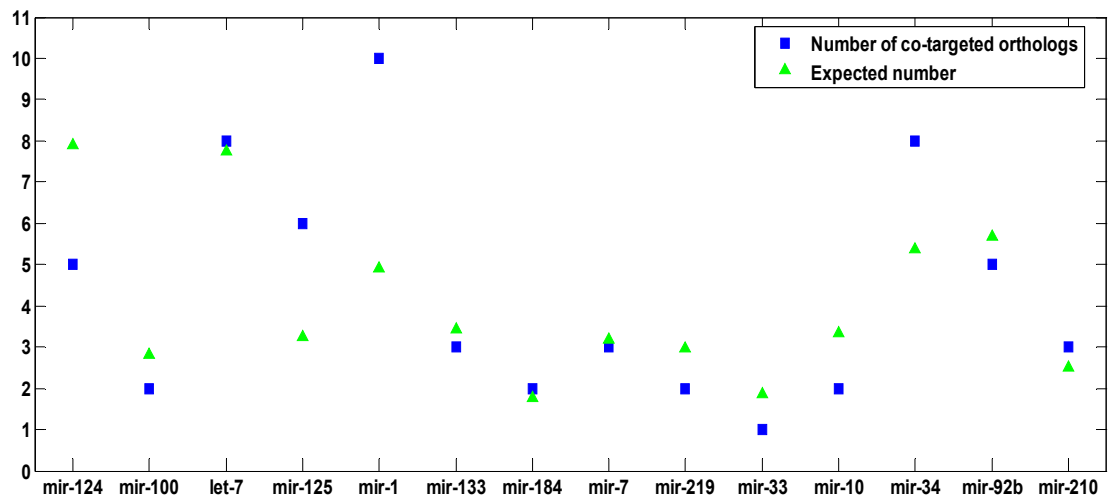

**Figure 1:** Number of genes targeted in both H. Sapiens and D. melanogaster. For each miRNA, the figure depicts (in blue) the number of orthologs of its H. sapiens targets which are also targeted in D. melanogatser. In green the expected number of targets is shown, if they are chosen at random from among the miRNA-targeted genes in D. melanogatser.

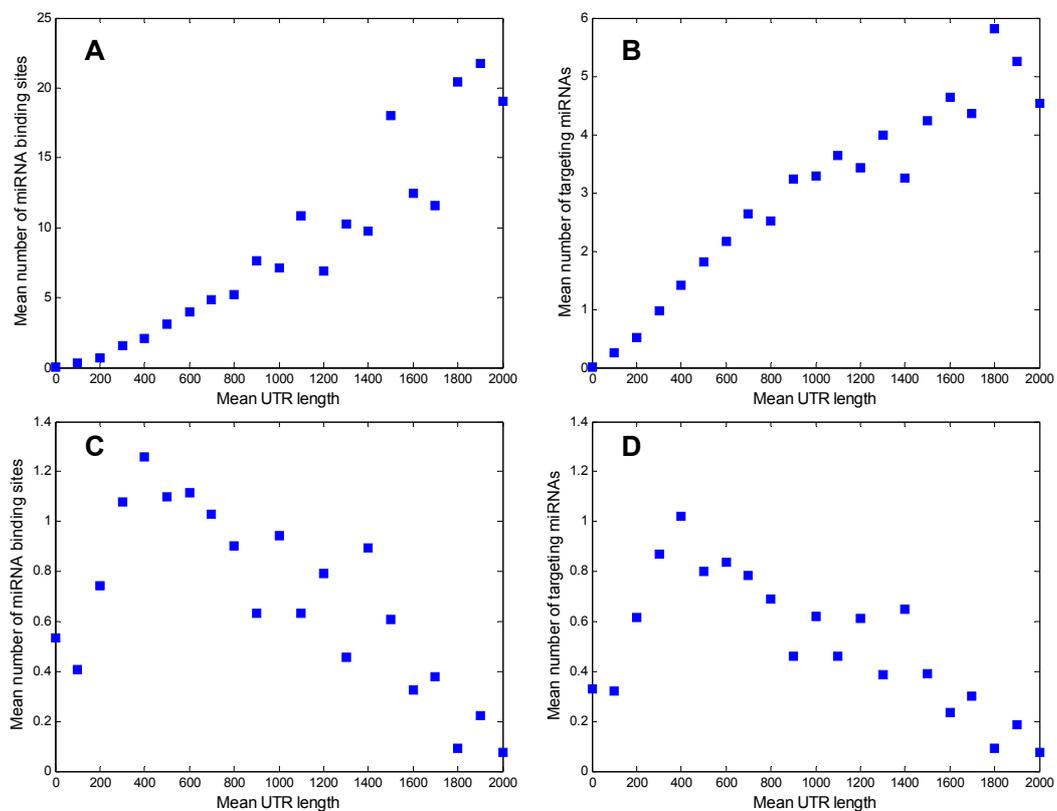

**Figure 2:** The effect of 3' UTR length on targeting by miRNA. **(A)** The mean number of miRNA binding sites, as a function of the 3' UTR length, based on PicTar target predictions. **(B)** The mean number of miRNA targeting a gene, as a function of the 3'

UTR length, based on PicTar target predictions. **(C)** As (A), based on miRanda target predictions. **(D)** As (B), based on miRanda target predictions.

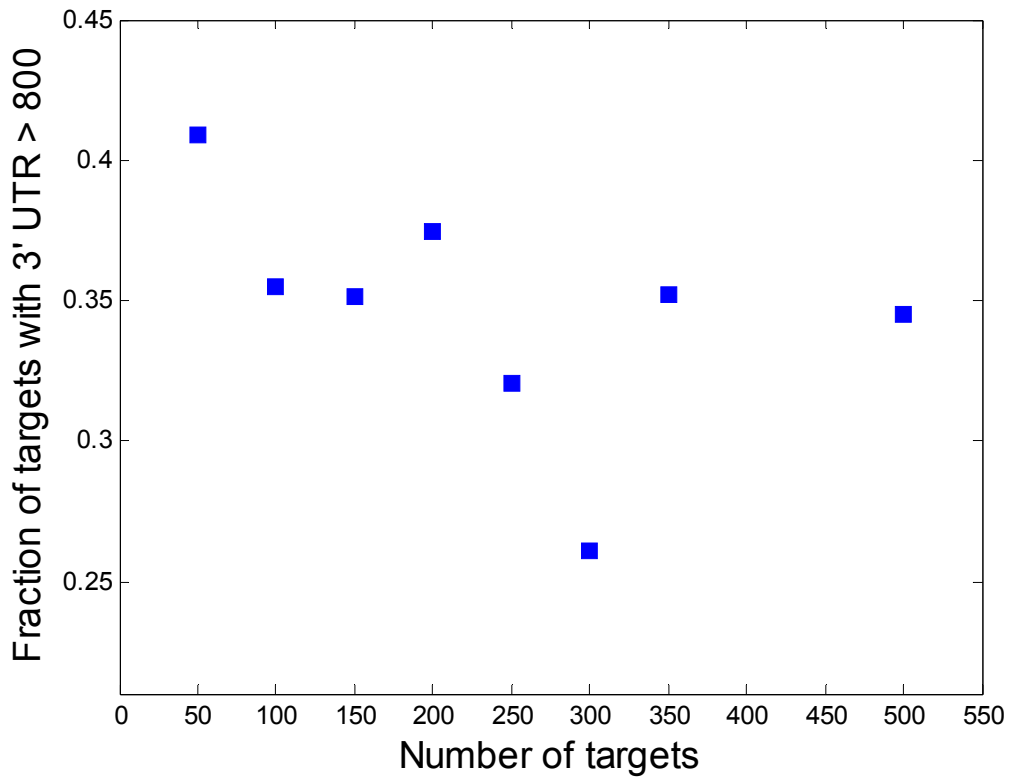

**Figure 3:** MicroRNAs with few targets prefer targets with long 3' UTR. All 78 miRNAs were binned into 10 bins according to the number of transcripts they target. The mean fraction of targets with a "long" (> 800bp) 3'UTR in each bin is depicted (PicTar target predictions).

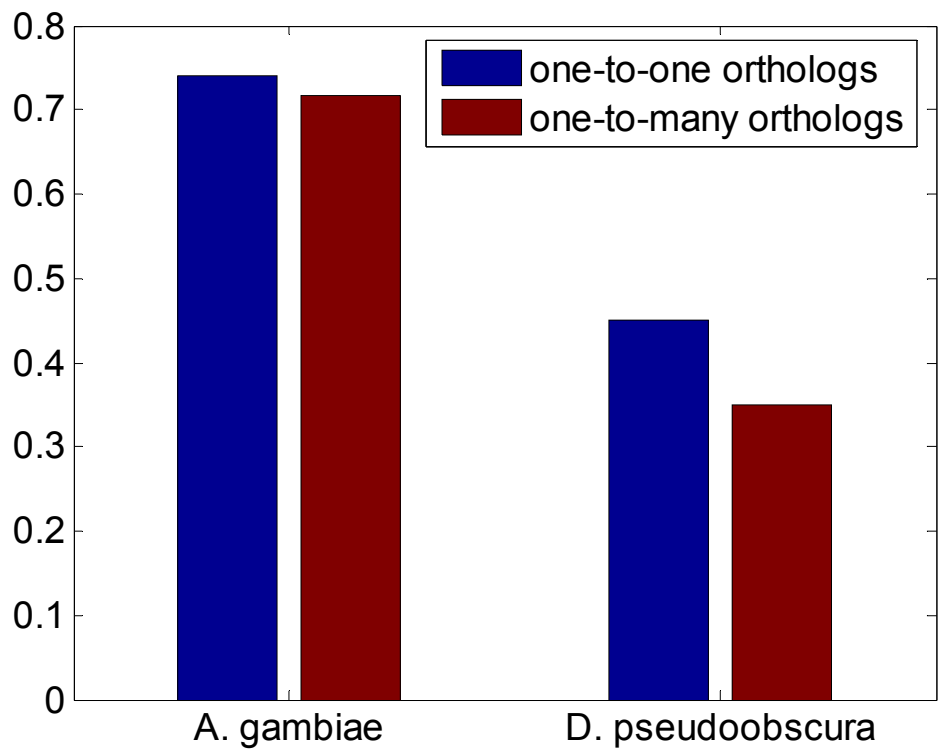

**Figure 4:** Mean number of miRNA targeted genes in A. gambiae and D. pseudoobscura, for genes with exactly one ortholog in D. melanogatser, and with multiple ortholgs (miRanda target predictions).